\documentclass[english,aps,prl,reprint,nofootinbib,superscriptaddress,longbibliography]{revtex4-2}

\usepackage[T1]{fontenc}
\usepackage[latin9]{inputenc}
\usepackage{amsmath}
\usepackage{amssymb}
\usepackage{graphicx}
\usepackage{wasysym}
\usepackage{color}
\usepackage{float}
\usepackage{placeins} 

\usepackage{hyperref}

\usepackage{xcolor}
\hypersetup{
	colorlinks,
	linkcolor={blue!70!black},
	citecolor={blue!70!black},
	urlcolor={blue!70!black}
}

\makeatletter
\def\clearfmfn{\let\@FMN@list\@empty}    
\@ifundefined{textcolor}{}
{%
 \definecolor{BLACK}{gray}{0}
 \definecolor{WHITE}{gray}{1}
 \definecolor{RED}{rgb}{1,0,0}
 \definecolor{GREEN}{rgb}{0,1,0}
 \definecolor{BLUE}{rgb}{0,0,1}
 \definecolor{CYAN}{cmyk}{1,0,0,0}
 \definecolor{MAGENTA}{cmyk}{0,1,0,0}
 \definecolor{YELLOW}{cmyk}{0,0,1,0}
 \definecolor{figa}{rgb}{0.7143, 0.7143, 0.7143}
 \definecolor{figb}{rgb}{0.4416, 0.7490, 0.4322}
 \definecolor{figc}{rgb}{0.3639, 0.5755, 0.7484}
 \definecolor{figd}{rgb}{0.9153, 0.2816, 0.2878}
 \definecolor{fige}{rgb}{1.0000, 0.5984, 0.2000}
 \definecolor{figf}{rgb}{0.6365, 0.3753, 0.6753}
}

\makeatother

\usepackage{pslatex}
\usepackage{babel}

\begin{document}

\title{Optical Trapping of a Polyatomic Molecule in an $\ell$-Type Parity Doublet State}

\author{Christian Hallas}
\email{christianhallas@g.harvard.edu}
\affiliation{Department of Physics, Harvard University, Cambridge, MA 02138, USA}
\affiliation{Harvard-MIT Center for Ultracold Atoms, Cambridge, MA 02138, USA}

\author{Nathaniel B. Vilas}
\affiliation{Department of Physics, Harvard University, Cambridge, MA 02138, USA}
\affiliation{Harvard-MIT Center for Ultracold Atoms, Cambridge, MA 02138, USA}

\author{Lo\"{i}c Anderegg}
\affiliation{Department of Physics, Harvard University, Cambridge, MA 02138, USA}
\affiliation{Harvard-MIT Center for Ultracold Atoms, Cambridge, MA 02138, USA}

\author{Paige Robichaud}
\affiliation{Department of Physics, Harvard University, Cambridge, MA 02138, USA}
\affiliation{Harvard-MIT Center for Ultracold Atoms, Cambridge, MA 02138, USA}

\author{Andrew Winnicki}
\affiliation{Department of Physics, Harvard University, Cambridge, MA 02138, USA}
\affiliation{Harvard-MIT Center for Ultracold Atoms, Cambridge, MA 02138, USA}

\author{Chaoqun Zhang}
\affiliation{Department of Chemistry, The Johns Hopkins University, Baltimore, MD 21218, USA}

\author{Lan Cheng}
\affiliation{Department of Chemistry, The Johns Hopkins University, Baltimore, MD 21218, USA}

\author{John M. Doyle}
\affiliation{Department of Physics, Harvard University, Cambridge, MA 02138, USA}
\affiliation{Harvard-MIT Center for Ultracold Atoms, Cambridge, MA 02138, USA}

\date{August 29, 2022}

\begin{abstract}
We report optical trapping of a polyatomic molecule, calcium monohydroxide (CaOH). CaOH molecules from a magneto-optical trap are sub-Doppler laser cooled to $20(3)~\mu\text{K}$ in free space and loaded into an optical dipole trap. We attain an in-trap molecule number density of $3(1) \times 10^9~\text{cm}^{-3}$ at a temperature of $57(8)~\mu$K. Trapped CaOH molecules are optically pumped into an excited vibrational bending mode, whose $\ell$-type parity doublet structure is a potential resource for a wide range of proposed quantum science applications with polyatomic molecules. We measure the spontaneous, radiative lifetime of this bending mode state to be $\sim$$0.7~\text{s}$.
\end{abstract}

\maketitle

Ultracold molecules are a promising platform for pursuing a large and diverse set of applications in quantum science \cite{carr2009cold}.
Recent experimental progress has led to ultracold samples of a broad range of heteronuclear diatomic species, produced via either direct laser cooling \cite{anderegg2018laser, caldwell2019deep, ding2020sub, Langin2021} or assembly of ultracold atoms \cite{zirbel2008, ni2008high, koppinger2014, Takekoshi2014, Park2015, voges2020, yu2021coherent}. These experiments have enabled new advances in physics such as the realization of degenerate gases of polar molecules \cite{de2019degenerate, schindewolf_evaporation_2022} and novel studies of ultracold collisions \cite{cheuk_2020, jurgilas_2021, anderegg_2021, gregory2021molecule, nichols_2022, Tobias2022}. Polyatomic molecules, as compared with diatomic molecules, have qualitatively richer internal level structures, which are expected to open up access to new, unique possibilities. 
Most prominently, polyatomic molecules generically possess low-lying, closely spaced levels with opposite parity. In small electric fields, these parity doublet structures result in long-lived, fully polarized quantum states, along with states that have zero first-order electric field sensitivity. These manifolds have been proposed for novel quantum simulation and computation platforms that have minimal field requirements and easily switchable interactions \cite{wei2011entanglement, wall2013simulating, wall2015quantum, wall2015realizing, yu2019scalable}, and for improved precision measurements of fundamental physics \cite{kozyryev2017precision, norrgardNuclear2019, hao_2020}. Ultracold polyatomic molecules would also facilitate new applications in beyond-standard-model searches \cite{kozyryev2021enhanced}, quantum chemistry \cite{krems2008cold, balakrishnan2016perspective, bohn2017cold, heazlewood2021towards}, and ultracold collisions \cite{augustovivcova2019ultracold}.

While offering new scientific capabilities, the internal complexity of polyatomic molecules also presents challenges to trapping and cooling. To date, only two species of polyatomic molecules, formaldehyde ($\text{H}_2\text{CO}$) and calcium monohydroxide ($\text{CaOH}$), have been trapped and cooled below $1 \text{ mK}$, using optoelectrical cooling in an electric trap and direct laser cooling in a magneto-optical trap (MOT), respectively \cite{prehn2016optoelectrical, vilas2022magneto}. 
Unlocking the many capabilities of polyatomic molecules will require techniques to reach even lower temperatures, as well as ways to attain long coherence times and high-fidelity quantum state control and readout. Optical trapping in particular has emerged as an important tool in ultracold physics for satisfying these criteria, as demonstrated by recent advances in quantum simulation and computation with neutral atoms \cite{bernien2017probing, levine2019parallel, madjarov2020high, schymik2020enhanced, scholl2021quantum} and trapping of individual diatomic molecules in optical tweezers \cite{cairncross_2021, Burchesky2021, zhang2022optical}. 
Developing methods to optically trap polyatomic molecules is highly desired in order to realize their full scientific and technological potential.

In this Letter, we report optical trapping of a polyatomic molecule, CaOH. Using a combination of two sub-Doppler laser cooling schemes, CaOH molecules are cooled to a temperature of 20(3) $\mu$K in free space and loaded into an optical dipole trap (ODT). Efficient state transfer of the trapped CaOH molecules to the $\widetilde{X}^2\Sigma^+(010)(N''=1^-)$ \cite{StateNotation} vibrational bending mode is accomplished via optical pumping. This state possesses an ``$\ell$-type'' parity doublet structure due to its near-degenerate vibrational angular momentum ($\ell$) levels \cite{QuantumNumberEll}. \textit{Ab initio} calculations for the spontaneous, radiative lifetime of the $\widetilde{X}^2\Sigma^+(010)(N''=1^-)$ state, $\tau_\text{rad}$, are performed and agree well with our experimentally measured value of $\tau_\text{rad} = 0.72_{-0.13}^{+0.25} \text{ s}$.

\begin{figure*}[t!]
    \centering
    \includegraphics{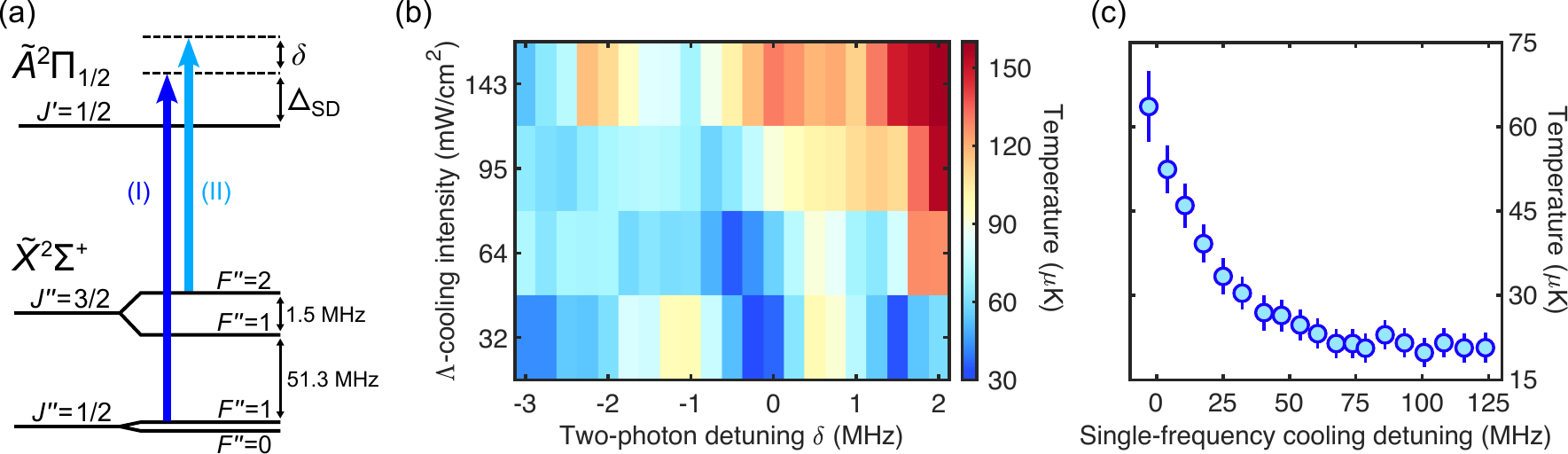}
    \caption{$\Lambda$-cooling and single frequency (SF) cooling of CaOH. (a) CaOH level structure and the two frequency components used for $\Lambda$-cooling (dark (I) and light blue (II) arrows), with a two-photon detuning of $\delta$ and single photon detuning of $\Delta_\text{SD}$. For SF cooling, only the bluer component (I) is present. (b) $\Lambda$-cooling temperature as a function of $\delta$ and intensity per cooling beam, $I_\text{SD}$. (c) Temperature of SF cooling as a function of $\Delta_\text{SD}$ at $I_\text{SD} = 64 \text{ mW/cm}^2$. Error bars include $1\sigma$ errors from the temperature fitting and systematic effects from imaging \cite{Supplemental}.}
    \label{fig:Fig1}
\end{figure*}

Our experiment starts with optical radiative slowing and magneto-optical trapping of CaOH molecules from a cryogenic buffer gas beam \cite{hutzler2012buffer}. Photon cycling is realized on the $\widetilde{X}^2\Sigma^+(000)(N'' = 1)\rightarrow \widetilde{A}^2\Pi_{1/2}(000)(J' = 1/2)$ electronic transition, which, during cooling, significantly populates eleven excited rovibrational states, each of which is optically repumped back into the cycling transition \cite{baum2021establishing, zhang2021accurate, vilas2022magneto}.
After the molecules are captured in the MOT, the MOT coils are turned off and the laser beams are retuned for sub-Doppler laser cooling and imaging of the molecular cloud. A detailed description of the apparatus and the repumping scheme is provided in Ref. \cite{vilas2022magneto}, where we reported cooling to a temperature of $\sim$$110~\mu\text{K}$. Here, we cool to lower temperatures by implementing two sub-Doppler cooling schemes in combination, $\Lambda$-cooling \cite{cheuk2018lambda, Langin2021} and single frequency (SF) cooling \cite{caldwell2019deep}. This approach was previously applied to diatomic molecules \cite{caldwell2019deep} and relies on the creation of zero-velocity dark states that lead to enhanced cooling through velocity-selective coherent population trapping (VSCPT). In $\Lambda$-cooling, counter-propagating lasers couple two ground states to a single excited state; a dark state at zero velocity occurs when the lasers are tuned to two-photon resonance \cite{aspect_1988}. In SF cooling, the zero-velocity dark states are created in the presence of a single frequency component of light that is blue-detuned from relevant ground state levels \cite{caldwell2019deep}.

Our $\Lambda$-cooling and SF cooling implementations for CaOH molecules are both based on creating a blue-detuned optical molasses that addresses the same $\widetilde{X}^2\Sigma^+ \rightarrow \widetilde{A}^2\Pi_{1/2}$ transition as the MOT. This is a ``type-II'' transition (i.e., $J'' \geq J'$), which exhibits strong sub-Doppler cooling (heating) at blue (red) detuning due to polarization gradient forces that arise from the existence of dark states in the ground state manifold \cite{Valentin_1992, boiron1995, sievers2015simultaneous, devlin2016three}. 
The experimental sequence for applying the molasses is as follows.
Starting with CaOH molecules loaded into the MOT as described in Ref. \cite{vilas2022magneto}, the molecular cloud is compressed in size by $\sim$$2\times$ by ramping the MOT gradient in a time period of 5 ms, from $8.3 \text{ G/cm}$ (its value during MOT loading) to $23.6 \text{ G/cm}$. 
The MOT laser beams, MOT coils, and MOT light polarization switching (which is used to remix magnetic dark states) are then turned off in a time period of $130$ $\mu$s. The two frequency components of the MOT beams---which separately address the spin-rotation components $\widetilde{X}^2\Sigma^+(N'' = 1, J'' = 1/2)$ and $\widetilde{X}^2\Sigma^+(N'' = 1, J'' = 3/2)$---are simultaneously tuned to the blue of the optical transition. The laser beams are then quickly turned back on to form the ``blue-detuned molasses,'' with detunings and intensities of the two frequency components chosen for implementing either $\Lambda$-cooling or SF cooling, as described below.

The blue-detuned molasses is configured for $\Lambda$-cooling by coupling two hyperfine levels in the $\widetilde{X}^2\Sigma^+(N'' = 1)$ manifold to the $\widetilde{A}^2\Pi_{1/2}(J'' = 1/2)$ excited state.  
To achieve this, the two frequency components of the molasses are tuned to nominally address the $\widetilde{X}^2\Sigma^+(J''=3/2, F''=2)$ and $\widetilde{X}^2\Sigma^+(J''=1/2)$ levels (Fig. \hyperref[fig:Fig1]{1(a)}), with polarizations $\sigma^-$ and $\sigma^+$, respectively. (The hyperfine level structure in the $\widetilde{X}^2\Sigma^+(J''=1/2)$ state is unresolved in our experiment, as $F''=0$ and $F''=1$ are split by only $\sim$$7.5 \text{ kHz}$ \cite{scurlock1993hyperfine}.) Both frequency components are blue-detuned by a common detuning $\Delta_\text{SD}$, while the component addressing $\widetilde{X}^2\Sigma^+(J''=3/2, F''=2)$ is further detuned by $\delta$, the two-photon detuning. We set $\Delta_\text{SD} = 12 \text{ MHz}$, informed by our previous sub-Doppler cooling results in Ref. \cite{vilas2022magneto}. The peak cooling intensity, $I_\text{SD}$, is divided between the two frequency components, whose intensities are $I_\text{SD}/3$ for $\widetilde{X}^2\Sigma^+(J''=3/2, F''=2)$ and $2I_\text{SD}/3$ for $\widetilde{X}^2\Sigma^+(J''=1/2)$. 

We investigate the dependence of $\Lambda$-cooling on both $\delta$ and $I_\text{SD}$, using ballistic expansion of the cooled molecules to measure their temperature after 1 ms of $\Lambda$-cooling (Fig. \hyperref[fig:Fig1]{1(b)}) \cite{Supplemental}. This cooling duration was chosen to be many times longer than the characteristic time of the cooling. The lowest measured temperature, $T_\text{min}=34(3)~\mu\text{K}$, occurs at $\delta \approx 0 \text{ MHz}$ and at the lowest cooling intensity used, $I_\text{SD} = 32 \text{ mW/cm}^2$. A second, slightly higher local temperature minimum is observed at $\delta \approx 1.5 \text{ MHz}$, which corresponds to the two-photon resonance for the $\Lambda$-system consisting of $\widetilde{X}^2\Sigma^+(J''=3/2, F''=1)$ and $\widetilde{X}^2\Sigma^+(J''=1/2)$. At higher intensities, the temperature is minimized at increasingly negative $\delta$. We attribute this behavior to ac Stark shifts of the $\Lambda$-coupled hyperfine levels due to the cooling light: for higher intensities, the levels move further apart, and smaller $\delta$ is required to satisfy the two-photon resonance condition.

\begin{figure}[b!]
    \centering
    \includegraphics{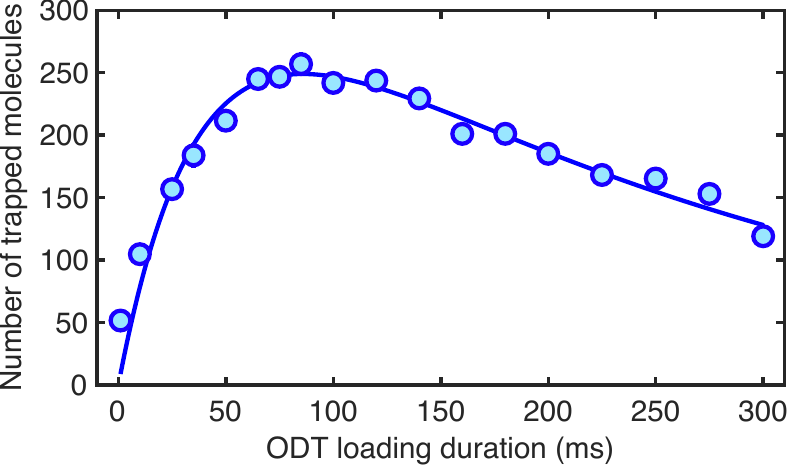}
    \caption{Number of CaOH molecules loaded into the optical dipole trap with single frequency (SF) cooling. The solid curve is a fit to a rate equation model with constant loading and loss rates. The loss rate is consistent with loss to vibrational dark states during SF cooling. Error bars are smaller than the data points.}
    \label{fig:Fig2}
\end{figure}

\begin{figure*}[t!]
    \centering
    \includegraphics{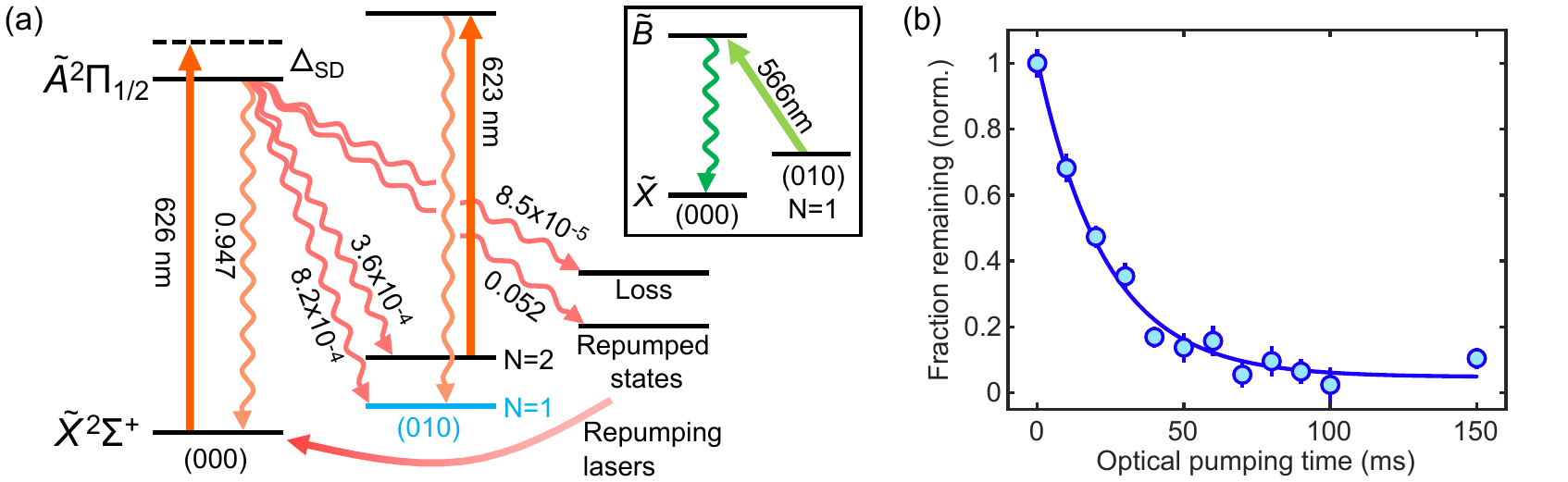}
    \caption{Populating the $\widetilde{X}^2\Sigma^+(010)(N''=1)$ vibrational bending mode. (a) The optical pumping scheme. The bending mode repumping laser is turned off while all other cycling and repumping lasers remain on. Vibrational repumping lasers for other states are shown schematically by the ``repumping'' arrow connecting back to $(000)$. Vibrational branching ratios are labeled for each decay channel. The main cooling laser is blue-detuned by a frequency $\Delta_\text{SD}$ to perform SF cooling of the molecules during population transfer. For more details on the optical cycling lasers used, see Ref. \cite{vilas2022magneto}. Inset: repumping transition driven to optically detect molecules in the bending mode. (b) Transfer into the bending mode versus optical pumping time. The solid curve is an exponential fit with a time constant of $\tau = 23.4(2.2)$ ms and an offset of 0.05(2).}
    \label{fig:Fig3}
\end{figure*}

SF cooling is implemented similarly to $\Lambda$-cooling, except that the frequency component used to address the $\widetilde{X}^2\Sigma^+(J''=3/2, F''=2)$ state is removed. We find that the capture velocity of SF cooling is too low to cool molecules directly from the MOT temperature of $\sim$$0.8~\text{mK}$. We therefore start by cooling the molecular cloud with $\Lambda$-cooling, then apply SF cooling. For this sequence of cooling, $\Lambda$-cooling with parameters of $\Delta_\text{SD} = 10~\text{MHz}$, $\delta = 0.25$ MHz, and $I_\text{SD} = 32 \text{ mW/cm}^2$ is applied for 2 ms, which cools the molecular cloud to $T\approx50$ $\mu$K. The molasses is next reconfigured for SF cooling in a period of 1 ms, during which time the molasses light is turned off. SF cooling is then applied for 5 ms, resulting in a minimum temperature of $T_\text{min} = 20(3)~\mu\text{K}$, realized when $I_\text{SD} = 64 \text{ mW/cm}^2$ and $\Delta_\text{SD} \gtrsim$ $70 \text{ MHz}$ (Fig. \hyperref[fig:Fig1]{1(c)}). The observed insensitivity of the cooling efficiency to detuning above a certain value ($\Delta_\text{SD} \approx 70 \text{ MHz}$) is similar to that observed in SF cooling of CaF molecules \cite{caldwell2019deep}. This insensitivity to detuning is potentially beneficial for cooling molecules into an ODT, where trap-induced light shifts might otherwise affect cooling efficiency.

CaOH molecules are loaded into the ODT using the blue-detuned molasses described above. The trapping potential is formed by a 1064 nm Gaussian laser beam with a power of 13.3 W and a waist size of $\sim$$25 \,\, \mu\text{m}$. \emph{Ab initio} calculations of the polarizability of the $\widetilde{X}^2\Sigma^+$ state predict a trap depth of $\sim$$600~\mu\text{K}$, which agrees well with the trap depth we measure using radial trap frequencies \cite{Supplemental}. The experimental sequence is as follows. After switching off the MOT, the ODT is turned on and a 1 ms pulse of $\Lambda$-cooling ($\Delta_\text{SD} = 12 \text{ MHz}$, $\delta = 0 \text{ MHz}$, $I_\text{SD} = 64 \text{ mW/cm}^2$) is used to cool the molecules to below the capture velocity of SF cooling. The molasses is then reconfigured for SF cooling ($\Delta_\text{SD} = 74 \text{ MHz}$, $I_\text{SD} = 64 \text{ mW/cm}^2$) to cool and load the molecules into the ODT. After loading the ODT, the molasses is turned off for $50 \text{ ms}$ to allow untrapped molecules to escape the field of view of the EMCCD camera used to image the molecules. Trapped molecules are then imaged by turning SF cooling on again for $100 \text{ ms}$ and collecting fluorescence decays from the $\widetilde{B}^2\Sigma^+$ state \cite{vilas2022magneto}. We find the optimal loading time (defined as the duration SF cooling is kept on concurrently with the ODT) for maximizing the number of molecules trapped in the ODT to be $\sim$$80~\text{ms}$, with $260(80)$ molecules loaded (Fig. \hyperref[fig:Fig2]{2}). Longer loading times decrease the number of loaded molecules due to loss to rovibrational dark states during SF cooling. An in-trap temperature of $T=57(8)~\mu\text{K}$ is determined by varying the ODT depth and observing the number of surviving molecules \cite{Supplemental}. We calculate a peak molecule number density of $3(1) \times 10^9 \text{ cm}^{-3}$ and a phase-space density of $9(5) \times 10^{-8}$ in the ODT \cite{Supplemental}. 

To study the $\widetilde{X}^2\Sigma^+(010)(N''=1^-)$ vibrational bending mode, we optically pump molecules into this state by simply turning off the corresponding repumping laser (Fig. \hyperref[fig:Fig3]{3(a)}). To quantify the transfer, we first load the ODT and image the trapped molecules with 50 ms of SF cooling, as described above. This first image serves as a normalization signal to account for shot-to-shot variations in the number of molecules loaded into the ODT. The molecules are then SF cooled for a variable time with the $\widetilde{X}^2\Sigma^+(010)(N''=1^-)$ repumping laser turned off. The vibrational branching ratio (VBR) from the optical cycle into this state is $\smash{v_\text{bend} = 8.2 \times 10^{-4}}$, while the VBR into dark rovibrational states is $\smash{v_\text{dark} \approx 8.5 \times 10^{-5}}$ \cite{vilas2022magneto}. We therefore expect the fraction of molecules pumped into the vibrational bending mode to be at most $\smash{v_\text{bend}/(v_\text{bend}+v_\text{dark}) \approx 91\%}$. The timescale for optical pumping, $\tau_\text{bend}$, corresponds to $\smash{1/v_\text{bend} \approx 1200}$ photons scattered at the SF cooling scattering rate, $R_\text{SF}$. At the measured scattering rate of $R_\text{SF} = 45 \times 10^3$ s$^{-1}$ \cite{Supplemental}, this results in a calculated optical pumping timescale of $\sim$$26$ ms. After optical pumping, the surviving ground-state molecules are imaged with the bending mode repumping laser turned off (Fig. \hyperref[fig:Fig3]{3(b)}). We observe $\tau_\text{bend} = 23.4(2.2)$ ms, which is consistent with the calculated value. A short repumping pulse on the $\widetilde{X}^2\Sigma^+(010)(N''=1) \rightarrow \widetilde{B}^2\Sigma^+(000)(N'=0)$ transition is used to recover molecules from the bending mode (Fig. \hyperref[fig:Fig3]{3(a), inset}). We observe 80(3)\% survival after two-way transfer into and out of the bending mode, normalized to the number of surviving molecules after an equal hold time in the $\widetilde{X}(000)$ ground state.

\begin{figure}[t!]
    \centering
    \includegraphics{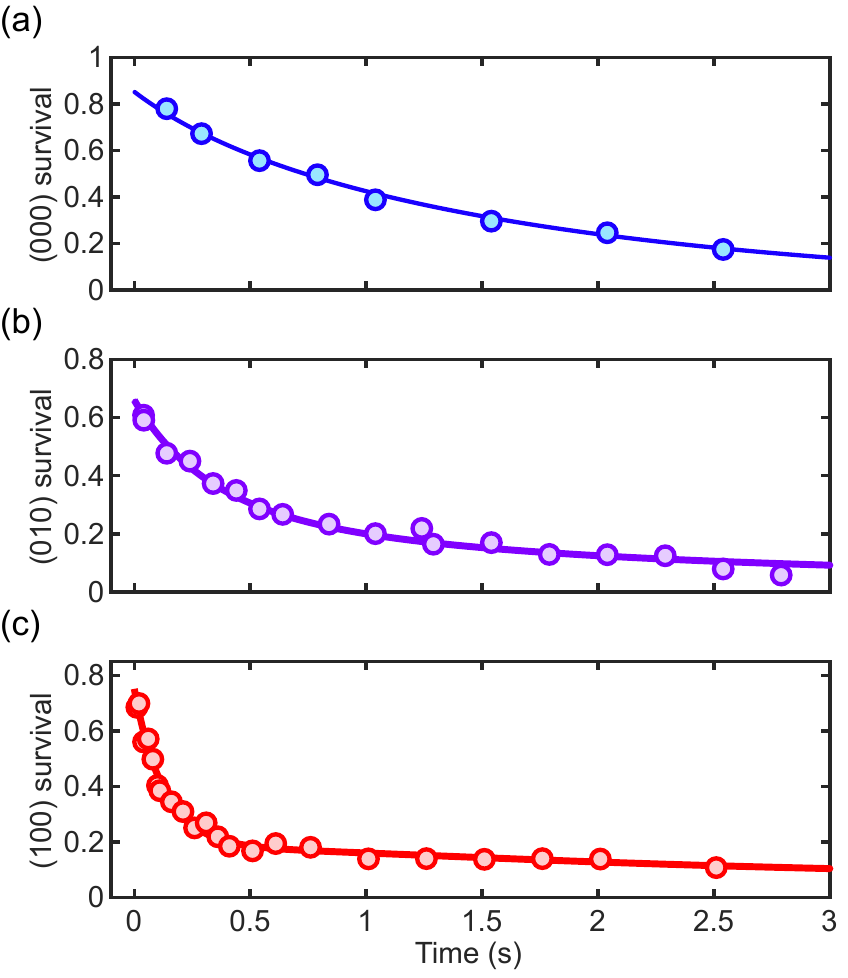}
    \caption{Lifetime of optically trapped CaOH molecules in the (a) $\widetilde{X}(000)$, (b) $\widetilde{X}(010)$, and (c) $\widetilde{X}(100)$ vibrational states. Data points are the fraction of molecules remaining in detectable rovibrational states after holding for a variable time with the majority of molecules prepared in the $(N''=1^-)$ level of the target vibrational state. Solid curves are fits to the rate equation model described in the text. Error bars are smaller than the data points.}
    \label{fig:Fig4}
\end{figure}

We measure the radiative lifetime of the $\widetilde{X}(010)$ vibrational bending mode, $\tau_\text{rad}$. The measurement sequence starts by optically pumping into the bending mode for 100 ms and then holding the molecules in the ODT for a variable amount of time with all cooling and repumping lasers off. The surviving detectable molecules, which include all molecules in the $(N''=1^-)$ level of the bending mode, along with the other vibrational states that are repumped in the optical cycling transition \cite{vilas2022magneto}, are then imaged for 50 ms. The surviving fraction, normalized to the number of molecules imaged before state transfer, is shown as a function of time in Fig. \hyperref[fig:Fig4]{4(b)}. Similar measurements are made for trapped CaOH molecules in the $\widetilde{X}(000)$ (Fig. \hyperref[fig:Fig4]{4(a)}) and $\widetilde{X}(100)$ (Fig. \hyperref[fig:Fig4]{4(c)}) vibrational states. The data for all three states are fit to a rate equation model (solid curves) that incorporates the effects of radiative decay and blackbody excitation between rovibrational states, as well as vacuum loss (see Ref. \cite{Supplemental} for more details). The rate equations capture the evolution of the vibrational populations over two distinct timescales: (i) initial vibrational thermalization due to radiative decay and blackbody excitation, and (ii) slow losses induced by blackbody excitation and vacuum loss. The fitted lifetime of the $\widetilde{X}(010)$ state, including all losses, is $0.36^{+0.11}_{-0.07} \text{ s}$ (68\% confidence interval), while the radiative lifetime determined from the fit is $\tau_\text{rad} = 0.72^{+0.25}_{-0.13} \text{ s}$. The measured lifetime of the $\widetilde{X}(000)$ state is $0.90^{+0.20}_{-0.16} \text{ s}$, dominated by the $1.3^{+0.3}_{-0.2} \text{ s}$ timescale for blackbody excitation to $\widetilde{X}(010)$ and $\widetilde{X}(100)$. The lifetime of $\widetilde{X}(100)$ is $0.14(2) \text{ s}$, dominated by the $0.19(3) \text{ s}$ lifetime for spontaneous, radiative decay to $\widetilde{X}(000)$. The fitted vacuum lifetime is $3.0^{+0.4}_{-0.7}$~s.

We perform {\it{ab initio}} calculations for the radiative lifetime of the $\widetilde{X}^2\Sigma^+(010)$ state of CaOH and compare with our experimental measurement. These calculations include anharmonic contributions and employ high-level treatments of electron correlation \cite{Supplemental}. The transition dipole moments between vibrational states of $\widetilde{X}^2\Sigma^+$ were calculated using an equation-of-motion coupled-cluster (EOM-CC) \cite{Stanton93a, EOMEA_Nooijen95} dipole moment surface and vibrational wave functions obtained from discrete variable representation \cite{Colbert92, zhang2021accurate} calculations on an EOM-CC potential energy surface. The calculated radiative lifetime of $\tau_\text{rad}=0.88$ s agrees well with the measured value. The same computational protocol produced similar and slightly longer $\widetilde{X}^2\Sigma^+(010)$ spontaneous lifetimes for SrOH and YbOH \cite{Supplemental}, suggesting that $\sim$$1 \text{ s}$ coherence times are attainable in several other molecules with similar vibrational structure to CaOH.

In summary, we have sub-Doppler cooled CaOH molecules to a temperature of $20(3)~\mu\text{K}$, loaded them into an optical dipole trap, and then transferred them into the long-lived $\widetilde{X}^2\Sigma^+(010)$ vibrational bending mode. The closely spaced $\ell$-type parity doublets in the bending mode allow >98\% polarization of the molecules at electric fields of $\sim$300 V/cm and should enable access to a variety of phenomena arising from dipolar interactions \cite{wall2013simulating,wall2015realizing,augustovivcova2019ultracold}, including applications that benefit from the existence of states with zero electric dipole moment alongside fully polarized, strongly interacting states \cite{yu2019scalable}. The spontaneous, radiative lifetime of the $\widetilde{X}^2\Sigma^+(010)$ bending mode is measured to be $\tau_\text{rad} \approx 0.7$ s. This suggests that experiments with coherence times near $\sim$$1 \text{ s}$ are achievable for vibrational bending modes in molecules of this type, provided that technical sources of loss can be addressed, e.g., by performing experiments in a mild cryogenic environment to suppress blackbody losses. Lifetimes of $\sim$$0.4$ s are currently attainable without apparatus upgrades. This also signals the potential of optically trapped molecules for spectroscopic measurements that would be difficult in a traditional beam-based apparatus, including their possible application to blackbody thermometry \cite{norrgard2022quantum}.

Optical trapping of polyatomic molecules opens paths to new precision measurements, including searches for dark matter \cite{kozyryev2021enhanced} and the electron electric dipole moment \cite{kozyryev2017precision}. In addition, the density of trapped CaOH molecules achieved here is sufficient for loading into optical tweezer arrays \cite{anderegg2019optical, zhang2022optical}, which would enable high-fidelity internal state manipulation of individual polyatomic molecules, controlled dipolar interactions between molecules in nearby traps, and studies of few-body interactions between molecules sharing a single trap \cite{cheuk_2020, anderegg_2021}. Higher molecule number densities could be achieved with improved ODT loading (e.g., through the use of repulsive bottle-shaped traps \cite{lu2022molecular}) and/or smaller trap geometries. At increased molecule number densities, studies of collisions \cite{augustovivcova2019ultracold}, ultracold chemistry \cite{bohn2017cold,heazlewood2021towards}, or bulk dipolar gases \cite{lahaye2009physics,baranov2012condensed} should be possible. Finally, this work suggests that optical trapping may be viable for several large classes of polyatomic molecules for which optical cycling is expected to be possible \cite{isaev2016polyatomic}, ranging from MOR type molecules \cite{kozyryev2016proposal} (of which several have already been laser cooled in one dimension \cite{kozyryev2017sisyphus, augenbraun2019laser, mitra2020direct}) to asymmetric molecules \cite{augenbraun2020molecular}, and possibly more complex species \cite{Ivanov2020, klos2020prospects, dickerson2021franck, zhu2022functionalization}. Several of these species (e.g., CaSH and CaOCH${}_3$) have closely-spaced parity doublet states that are expected to have even longer radiative lifetimes ($>$$10~\text{s}$) than the $\widetilde{X}^2\Sigma^+(010)$ state in CaOH.

This work was supported by the AFOSR and the NSF. NBV acknowledges support from the NDSEG fellowship, LA from the HQI, and PR from the NSF GRFP. The computational work at Johns Hopkins University was supported by the NSF under Grant No. PHY-2011794.

\bibliographystyle{apsrev4-2}

\clearpage
\newpage

\onecolumngrid
\begin{center}
	\textbf{\large Supplemental Material}
\end{center}

\setcounter{figure}{0}

\makeatletter 
\renewcommand{\thefigure}{S\@arabic\c@figure}
\makeatother

\section{Sub-Doppler cooling temperature measurements}
The temperatures of sub-Doppler cooled CaOH molecules are determined from the standard ballistic expansion method. After the sub-Doppler cooling light has turned off, the cooled molecular cloud is allowed to expand ballistically for a variable amount of time, $\tau_\text{exp}$. An EMCCD camera image of the molecular cloud is then taken by applying resonant light on the $\widetilde{X}^2\Sigma^+(000)(N'' = 1) \rightarrow \widetilde{A}^2\Pi_{1/2}(000)(J' = 1/2)$ cycling transition with the MOT beams for 1 ms and collecting fluorescence decays from the $\widetilde{B}^2\Sigma^+$ state \cite{vilas2022magneto}.  Radial and axial Gaussian widths, $\sigma_\rho$ and $\sigma_z$, respectively, of the imaged cloud are fitted from a 2D Gaussian model and used to extract radial and axial temperatures from fits to $\sigma_{\rho, z}^2 = \sigma_{0, \rho, z}^2 + k_B T_{\rho, z} \tau_\text{exp}^2 / m$, where $\sigma_{0, \rho, z}$ are the initial widths of the molecular cloud, $k_B$ is the Boltzmann constant, $T_\rho$ and $T_z$ are the radial and axial temperatures, and $m$ is the mass of CaOH. The temperature of the molecules is reported as the resulting geometric mean \smash{$T = T_\rho^{2/3} T_z^{1/3}$}. 

Resonant imaging of the molecular cloud causes heating that could lead to additional expansion of the molecular cloud. The imaging time of 1 ms was chosen to be short compared to the expansion times to minimize this effect, and it was verified by comparing to images with shorter exposures of 0.5 ms that the imaging did not significantly affect the size of the molecular cloud. The magnification of the imaging system was calibrated to 1.16(3) by fitting the vertical position of a falling molecular cloud to $y = a \tau_\text{exp}^2 / 2$ and comparing the fitted acceleration $a$ to the known acceleration due to gravity. The errors reported for the sub-Doppler cooling temperatures include both the statistical errors from the fitting procedure described above along with the error introduced from the uncertainty in the magnification.

\section{Scattering rate for SF cooling in the trap} \label{scattering_rate}
\begin{figure}[b]
	\centering
	\includegraphics{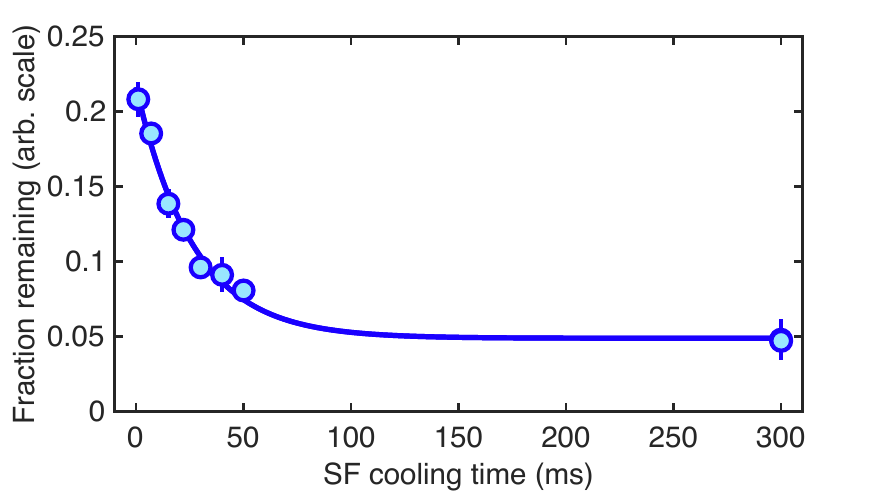}
	\caption{Molecule loss versus cooling time for SF cooling in the ODT, with repumping lasers removed so that the molecules scatter only 1200 photons before $1/e$ loss occurs. The fit is to an exponential decay with a time constant $\tau_\text{scatt} = 26.7(2.3)$ ms and an offset of $0.049(4)$.}
	\label{fig:scatteringrate}
\end{figure}

To measure the scattering rate for SF cooling in the ODT (which is useful for characterizing the \emph{in situ} imaging and optical pumping into the bending mode), we load molecules into the ODT and image for 60 ms for normalization of the initial number. The SF imaging/cooling parameters are $I_\text{SD}=64~\text{mW/cm}^2$ and $\Delta_\text{SD} = 74$ MHz. We then cool the molecules for a variable time with the same SF cooling parameters, but with several repumping lasers removed so that only 1200 photons are scattered before $1/e$ loss of molecules occurs. The cycling and repumping lasers used for the measurement address $\widetilde{X}(000)$, $(100)$, $(02^00)$, $(200)$, $(02^20)$, and $(010)(N=1^-)$ only. After cooling, the surviving molecules are imaged (with the high-level repumpers still turned off), and the signal is divided by the number in the first image to determine the survival fraction. A fit to an exponential function with an offset is shown in Fig. \ref{fig:scatteringrate}. The offset arises primarily from molecules which are transferred from dark to bright rovibrational states due to vibrational thermalization between cooling and imaging. The fitted exponential decay time is $\tau_\text{scatt} = 26.7(2.3)$ ms, corresponding to a scattering rate $R_\text{SF} = 45(4) \times 10^3$ s$^{-1}$.

\section{Number of molecules trapped in ODT}
The captured images of trapped CaOH molecules are converted to a number of trapped molecules by characterizing the three factors included in this conversion: (i) the number of photons scattered per molecule during imaging, (ii) the detection efficiency of the camera, and (iii) the collection efficiency of the imaging optics. The number of photons scattered can be determined from the photon scattering rate at the SF cooling parameters used for imaging, which is measured to be $45(4) \times 10^3$ s$^{-1}$ (see Sec. \ref{scattering_rate}). Since the detected fluorescence is due to decay from the $\widetilde{B} \rightarrow \widetilde{X}$ transition via repumping through $\widetilde{B}$, only one in every $\sim$18 photons scattered during optical cycling is detectable \cite{vilas2022magneto}. 
The detection efficiency of the camera was calibrated by imaging a known amount of light with similar intensity and wavelength (to within 1 nm) to our usual signal from the molecules. Finally, the collection efficiency of the imaging system, including transmission losses through the optics, is determined to be $0.20(7)\%$ from a combination of calibration measurements and estimates of the numerical aperture. 

\section{Trap depth of ODT inferred from radial trap frequency measurements} \label{trap_depth}
\begin{figure}[b]
	\centering
	\includegraphics{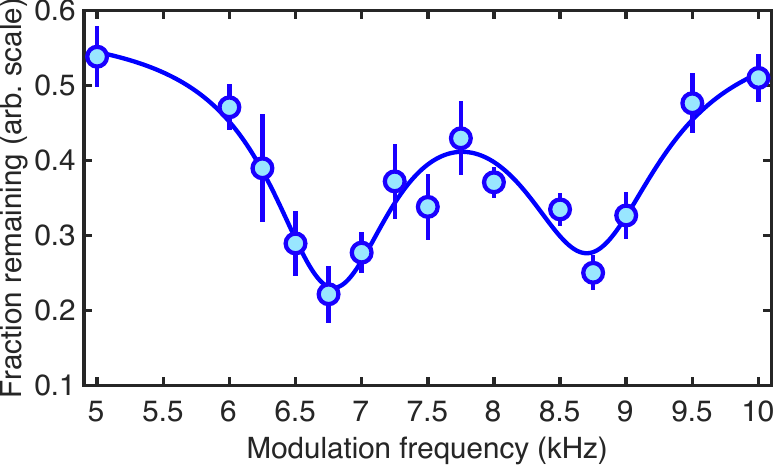}
	\caption{Molecule loss versus modulation frequency of the trap depth of the ODT. The fit is to two Lorentzian functions with centers located at $6.8(1) \text{ kHz}$ and $8.7(2) \text{ kHz}$ and an offset of 0.58(8).}
	\label{fig:trap_frequency}
\end{figure}

The radial trap frequencies of the ODT are determined by the parametric heating method \cite{grimm2000optical}. Following loading of the ODT, a first image is taken with 50 ms of SF cooling. The power of the ODT is then amplitude modulated from the nominal power (13.3 W) with a variable modulation frequency and a modulation depth of $\sim${}$29\%$ for a total of 500 modulation cycles. This modulation sequence is implemented by modulating the input voltage to an acousto-optic modulator (AOM) installed in the optics setup of the ODT. A second 50 ms image is taken following the modulation sequence, with a short wait time inserted between the end of the modulation and the second image to ensure that the first and second images are taken 120 ms apart regardless of the duration of the modulation sequence. Molecules are held in the ODT at nominal power during this wait time. Loss of molecules from the ODT due to parametric heating caused by the modulation appears as a decrease in the fraction of molecules left in the second image as compared to the first image (Fig. \ref{fig:trap_frequency}). Two resonant loss features are observed at $6.8(1)$ kHz and $8.7(2)$ kHz, from which we identify radial trap frequencies $\omega_x = 2\pi \times 6.8(1) / 2 = 2\pi \times 3.4(1)$ kHz and $ \omega_y = 2\pi \times 8.7(2)/2 = 2\pi \times 4.4(1)$ kHz. The axial trap frequency was not measured.

The two distinct radial trap frequencies indicate that the Gaussian beam forming the ODT has anisotropy in the transverse plane. To account for this in the estimate of the trap depth of the ODT, we model the trapping potential as that due to a Gaussian beam profile with two separate waist sizes $w_{0,x}$ and $w_{0,y}$ \cite{simon_1983},
\begin{equation}
	U_\text{dip}(x, y, z) 
	= - U_\text{0} \frac{w_{0,x} w_{0,y}}{w_x(z) w_y(z)}
	\exp\left\{
	- 2
	\left( 
	\frac{x^2}{w_x(z)^2} 
	+ \frac{y^2}{w_y(z)^2} 
	\right)
	\right\},
\end{equation}
where $U_\text{0}$ is the trap depth and \smash{$w_{\alpha}(z) = w_{0,\alpha} [1 + (z / z_{R,\alpha})^2]$}, where $\alpha \in \{x,y\}$. Here $z_{R,\alpha} = \pi w_{0,\alpha} / \lambda$ are the Rayleigh lengths associated with each axis, and $\lambda = 1064 \text{ nm}$ is the wavelength of the ODT. At low temperatures compared to $U_\text{0}$ (which is the case for our ODT, see Sec. \ref{ODT_temp_density}), the trap density is concentrated close to the origin, and a Taylor expansion to second order is then valid:
\begin{align} \label{eq:taylor}
	U_\text{dip}(x, y, z) 
	&\approx 
	U_\text{0} \left(
	\frac{2 x^2}{w_x^2} + \frac{2 y^2}{w_y^2} + 
	\frac{1}{2} \left(\frac{1}{z_{R,x}^2} + \frac{1}{z_{R,y}^2}\right) z^2 - 1
	\right),
\end{align}
Noting that the potential resembles that of a harmonic trap, we can similarly write it in the familiar form
\begin{align} \label{eq:harmonic}
	U_\text{dip}(x, y, z) 
	&\approx 
	\frac{1}{2} m 
	\left(
	\omega_x^2 x^2 + \omega_y^2 y^2 + \omega_z^2 z^2
	\right)
	- U_\text{0},
\end{align}
where $\omega_z$ is the axial trap frequency. Comparing equations \ref{eq:taylor} and \ref{eq:harmonic}, we find that
\begin{equation} \label{eq:trap_depth_eq}
	U_\text{0} = \frac{m w_{\alpha}^2 \omega_{\alpha}^2}{4}.
\end{equation}
The radial waists are measured to be $w_x = 27.5$ $\mu\text{m}$ and $w_y = 23.3$ $\mu\text{m}$ by imaging the focused ODT onto a CCD camera and fitting the image to a 2D Gaussian. Combined with the trap frequencies reported above, the trap depth may be calculated from eqn. \ref{eq:trap_depth_eq} using either $w_x$ and $\omega_x$ or $w_y$ and $\omega_y$, resulting in two separate measurements of the trap depth of $\sim$$592~\mu\text{K}$ and $\sim$$704~\mu\text{K}$. The $\sim$15\% disagreement between these two values may be due to uncharacterized errors in the waist and/or frequency measurements or possible deviations of the trapping potential from an asymmetric Gaussian beam profile, for example, due to astigmatism, which is not modeled. For density measurements (see Sec. \ref{ODT_temp_density}), we use the geometric mean, i.e., $U_0 = m \overline{w}^2 \overline{\omega}^2 / 4 = 645(87)~\mu\text{K}$, where $\overline{w} = (w_x w_y)^{1/2}$ and $\overline{\omega} = (\omega_x \omega_y)^{1/2}$.

\begin{figure}[b]
	\centering
	\includegraphics{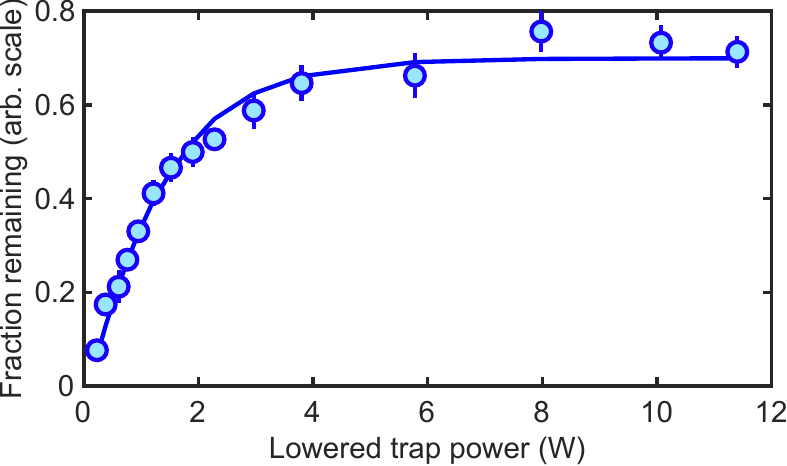}
	\caption{Molecule loss versus lowered trap depth. The solid curve is a Monte Carlo simulation of losses of trapped molecules, initialized with a temperature $57\,\,\mu\text{K}$, during the experimental sequence corresponding to the data shown.}
	\label{fig:shaving_temperature}
\end{figure}

\section{Temperature, density, and phase-space density measurements in the ODT} \label{ODT_temp_density}
The temperature of molecules trapped in the ODT is determined by studying the loss of molecules from the trap when the trap depth is lowered. A first image of the trapped molecules is taken with 50 ms of SF cooling to measure the initial number of molecules loaded into the ODT. After a 40 ms wait time following this first image, the trap depth is lowered by decreasing the ODT power to a variable fraction of the initial power (13.3 W). The ODT power is controlled by an AOM (see Sec. \ref{trap_depth}) and is changed with a response time of $\sim$10 $\mu$s. The ODT is kept at this lower power for 30 ms before increasing the power back to its initial value. Following a 30 ms wait, a second 50 ms image is taken and the surviving fraction of molecules is determined by comparison to the first image (Fig. \ref{fig:shaving_temperature}). In order to extract the in-trap temperature from the experimentally observed loss fraction versus trap depth, the trap losses are modelled by a Monte Carlo simulation of trapped CaOH molecules undergoing the experimental sequence. In the simulation, molecules are initialized at a given temperature with velocities following a Maxwell-Boltzmann distribution and positions sampled from the spatial density distribution assuming a harmonic trap potential (eqn. \ref{eq:harmonic}). The molecules undergo classical motion due to the force from the trapping potential of the ODT as well as gravity, $\mathbf{F} = - \nabla U_\text{dip}(t) - g \hat{\mathbf{y}}$, where $g$ is the acceleration due to gravity and the time dependence of $U_\text{dip}(t)$ is dictated by the experimental sequence. At the end of the simulation, molecules with final positions $(x,y,z)$ such that $|x| > 10 w_x$, $|y| > 10 w_y$, or $|z| > 10 \, \text{max}(z_{R,x}, z_{R,y})$ are considered ``lost''. The simulation is insensitive to the exact choice of these bounds since, due to gravity, untrapped molecules typically fall vertically much below $-10 w_y$ before the second image is taken. We run the simulation for a range of initial temperatures and, with experimental data obtained from the experimental sequence above, compute the $\chi^2$ error for each temperature. The resulting $\chi^2$ curve versus temperature follows a second-order polynomial about the minimum \cite{BevingtonRobinson}, which we use to determine an in-trap temperature of $T_\text{in-trap} = 57(8)\,\,\mu\text{K}$. The error in the temperature is dominated by the error in the determination of the trap depth of the ODT (see Sec. \ref{trap_depth}).

With the temperature estimate above, we estimate the peak density of the ODT as 
\begin{equation}
	n_0 = 
	N
	\left(
	\frac{m (\omega_x \omega_y \omega_z)^{1/3}}{2\pi k_B T_\text{in-trap}}
	\right)^{3/2}
	=
	3(1) \times 10^9,
\end{equation}
where $\omega_z = [U_\text{0} (1/z_{R,x}^2 + 1/z_{R,y}^2) / m]^{1/2} = 235(16) \text{ Hz}$ is determined by comparing equations \ref{eq:taylor} and \ref{eq:harmonic} and using the trap depth $U_0$ determined in Sec. \ref{trap_depth} along with $z_{R,x} = 2.2 \text{ mm}$ and $z_{R,y} = 1.6 \text{ mm}$. The corresponding peak phase-space density is
\begin{equation}
	\rho_0 = n_0 \lambda_\text{dB}^3 = 9(5) \times 10^{-8},
\end{equation}
where $\lambda_\text{dB} = (2\pi \hbar^2 / m k_B T_\text{in-trap})^{1/2} = 3.0(2) \times 10^{-8}$.

\section{Rate equation model for vibrational lifetimes}

We model the internal state dynamics of CaOH molecules trapped in an ODT using a set of rate equations that capture the effects of radiative decay and blackbody excitation between vibrational manifolds within the $\widetilde{X}^2\Sigma^+$ electronic ground state. We consider rovibronic states described by the quantum numbers $\lvert v_1, v_2, \ell, N, J, p\rangle$, where $v_1$ and $v_2$ are the vibrational quantum numbers of the Ca--O stretch and Ca--O--H bending modes, $\ell$ is the magnitude of the vibrational angular momentum projected onto the molecular axis, $N$ is the rotational quantum number, $J$ is the total electronic angular momentum, and $p$ is the parity. We neglect the O--H stretching mode because its frequency is far from the peak of the blackbody spectrum at room temperature.  All states with $\ell=0$ have $\Sigma^+$ symmetry and their parity is $p=(-1)^N$. For states with $\ell \neq 0$, each rotational manifold has two opposite parity levels, $\lvert \ell, N, \pm\rangle = 2^{-1/2}\left\{\lvert \ell, N\rangle \pm (-1)^{N-\ell}\lvert -\ell, N\rangle\right\}$. Hyperfine structure is unresolved by the lasers and is therefore omitted. In the simulations, we include all vibrational levels up through $v_1=2$ and $v_2=2$, and all rotational levels up through $N=5$. The populations of the highest states included in each rotational and vibrational ladder are typically $<$$5\%$ over the experimental timescale ($t\lesssim5$ s), implying that inclusion of higher-energy states is unnecessary.

The spontaneous decay rate, $A_{ij}$, and the blackbody excitation rate, $R_{ij}$, from initial state $i$ to final state $j$ are \cite{bernath2005spectra}
\begin{align}
	A_{ij} &= \frac{\omega_{ij}^3}{3\pi\varepsilon_0\hbar c^3 (2J_i +1)}S_{ij}
	\label{eqn:spont},
	\\
	R_{ij} &= \frac{1}{6\varepsilon_0\hbar^2(2J_i+1)}\frac{2\hbar \omega_{ij}^3}{\pi c^3}\frac{1}{e^{\hbar\omega_{ij}/k_BT}-1}S_{ij}
	\label{eqn:BBR},
\end{align}
where $S_{ij} = \left\lvert \langle i \lvert \mu \lvert j \rangle \right\lvert^2$ is the rovibrational transition strength, $J_i$ is the angular momentum quantum number of the initial state, $\omega_{ij}$ is the transition frequency, and $T$ is the temperature of the environment. In the experiment, the MOT coils are cooled to $\sim$0$^\circ$C to minimize outgassing and improve the vacuum pressure. The solid angle of the coils is $0.57\times 4\pi$ sr, so in the model we include two terms in eqn. \ref{eqn:BBR} such that 57\% of the blackbody radiation seen by the molecules is at 273 K while the remaining 43\% is at room temperature (295 K).

The rovibrational transition strengths can be separated into vibrational and rotational components:
\begin{align}
	S_{ij} &= \left\lvert \langle v_{1i}, v_{2i}, \ell_i, N_i, J_i, p_i \lvert \mu \lvert v_{1j}, v_{2j}, \ell_j, N_j, J_j, p_j \rangle \right\lvert^2 \nonumber \\
	&= \left\lvert \langle v_{1i}, v_{2i}, \ell_i \lvert \mu \lvert v_{1j}, v_{2j}, \ell_j \rangle \right\lvert^2 \left\lvert \langle \ell_i, N_i, J_i, p_i \lvert \mu \lvert \ell_j, N_j, J_j, p_j \rangle \right\lvert^2 \nonumber \\
	&\equiv S^\text{vib}_{ij} S^\text{rot}_{ij},
\end{align}
where $S^\text{rot}_{ij}$ and $S^\text{vib}_{ij}$ are the rotational and vibrational line strengths, respectively. The rotational line strengths, or H\"onl-London factors, are \cite{hansson2005comment, hirota2012high}
\begin{align}
	S^\text{rot}_{ij} = \delta_{p_i,-p_j}(1+\delta_{\ell_i 0} + \delta_{\ell_j 0} - 2 \delta_{\ell_i0}\delta_{\ell_j0})&(2N_i+1)(2N_j+1)
	\begin{pmatrix}
		N_i & 1 & N_j \\
		-\ell_i & \ell_i-\ell_j & \ell_j
	\end{pmatrix}
	^2 (2J_i+1)(2J_j+1)
	\begin{Bmatrix}
		N_j & J_j & S \\
		J_i & N_i & 1
	\end{Bmatrix}
	^2,
	\label{eqn:hlvib}
\end{align}
where we emphasize that $\ell$ is always a positive number since the parity is defined for each state.

For the vibrational line strengths, we make the ``double harmonic'' approximation \cite{bernath2005spectra}, wherein the vibrational potential is assumed to be that of a perfect harmonic oscillator, and the electric dipole moment is assumed to be a linear function of the internuclear spacing near the equilibrium geometry of the molecule. Within this approximation, we expand the vibrational line strength as follows:
\begin{align}
	\langle v_{1i}, v_{2i}, \ell_i \lvert \mu \lvert v_{1j}, v_{2j}, \ell_j \rangle \approx \left|\frac{d\vec{\mu}}{dQ_1}\right|_{Q_{1,\text{eq}}} \langle v_{1i}\lvert Q_1 \lvert v_{1j}\rangle + \left|\frac{d\vec{\mu}}{dQ_2}\right|_{Q_{2,\text{eq}}} \langle v_{2i}, \ell_i \lvert Q_2 \lvert v_{2j}, \ell_j \rangle,
\end{align}
where $Q_k$ is the normal coordinate of the $k$th vibrational mode, and $Q_{k,\text{eq}}$ is its equilibrium value. Within the harmonic approximation, we can use 1D and 2D harmonic oscillator algebra to write the matrix elements for the stretching and bending modes, respectively:
\begin{align}
	\left\lvert\langle v_1+1|Q_1|v_1\rangle\right\lvert^2 &= \frac{v_1+1}{2}, \\
	\left\lvert\langle v_2+1, \ell\pm1|Q_2|v_2,\ell\rangle\right\lvert^2 &= \frac{1}{4}(1+ \delta_{\ell,0} + \delta_{\ell\pm1,0}-\delta_{\ell,0}\delta_{\ell\pm1,0})\left(\frac{v_2\pm\ell}{2}+1\right).
\end{align}
The advantage of the double-harmonic approximation is that it allows every vibrational transition moment to be expressed in terms of just two parameters, the dipole moment derivatives along each of the normal coordinates. This makes fitting the rate equations to the lifetime measurements tractable. By comparison, an anharmonic model introduces additional fit parameters, many of which would likely be underdetermined. Because the majority of the dynamics considered in this work occur in vibrational levels near the bottom of the molecular potential, the harmonic approximation is expected to be reasonably appropriate.  Nonetheless, the fitted values of $|d\vec{\mu}/dQ|$ should be interpreted as ``effective'' parameters that include the contribution of anharmonic effects on the measured $(000)$, $(010)$, and $(100)$ lifetimes. These contributions are expected to be relatively small: the \textit{ab initio} calculations performed in Sec. \ref{ab_initio} indicate that anharmonicity in the molecular potential is expected to have a $\sim$10-20\% effect on the vibrational lifetimes.

The internal state dynamics under the influence of blackbody excitation, spontaneous emission, and vacuum losses are determined by a set of rate equations for the population, $n_i$, of rovibrational state $i$:
\begin{equation}
	\frac{dn_i}{dt} = - \sum_j R_{ij}n_i - \sum_{j<i} A_{ij}n_i + \sum_j R_{ji}n_j + \sum_{j>i} A_{ji}n_j - \frac{n_i}{\tau_\text{vac}}
	\label{eqn:rateeqn},
\end{equation}
where $\sum_{j<i}$ implies a sum over all states lower in energy than state $i$, and $\tau_\text{vac}$ is the (state-independent) vacuum lifetime. To preserve normalization of the population vector, a ``lost'' population is also included according to
\begin{equation}
	\frac{dn_\text{loss}}{dt} = \sum_i \frac{n_i}{\tau_\text{vac}}.
\end{equation}

The rate equations are solved by numerically integrating eqn. \ref{eqn:rateeqn} (sometimes with additional terms described below) applied to a population vector including all states with $v_\text{stretch} \leq 2$, $v_\text{bend} \leq 2$, $N\leq 5$ along with the ``lost'' population $n_\text{loss}$, which encompasses both vacuum loss and loss to dark states during optical cycling (neither of which is recoverable). The sequence is described as follows, and is chosen to exactly match the experimental protocol unless otherwise specified. The simulation is initialized with all molecules in the untrapped state, $n_\text{loss}(t=0) = 1$. This population is pumped/loaded into the ODT, in $\widetilde{X}^2\Sigma^+(000)(N=1, J=3/2)$, with a characteristic rate of $R_\text{load}=1/(48.5 ~\text{ms})$, fit from the experimental data (Fig. 2 of the main text). Here and below, it is assumed that population accumulates entirely in the $\widetilde{X}^2\Sigma^+(000)(N=1, J=3/2)$ level when the optical cycling light is on. This is because this state is the furthest-detuned ground-state level during SF cooling; additionally, the VSCPT dark state consists of hyperfine levels within this manifold \cite{caldwell2019deep}. Whenever SF cooling light is on, an additional term is added to the rate equations which pumps population from $\widetilde{X}^2\Sigma^+(000)(N=1, J=1/2)$ to $\widetilde{X}^2\Sigma^+(000)(N=1, J=3/2)$ at a rate corresponding to the SF scattering rate \cite{tarbutt2015magneto}.

As soon as the molecules are trapped, their internal states begin evolving according to eqn. \ref{eqn:rateeqn}. Prior to ODT loading, these dynamics are suppressed because any molecule which is blackbody-excited out of the cycling scheme will not be cooled/loaded into the ODT. After the ODT loading time of 100 ms, the loading is turned off and the remaining molecules evolve through each step in the experimental sequence. First, the molecules are held in the ODT and propagate according to eqn. \ref{eqn:rateeqn} for 30 ms. Next, the molecules undergo 50 ms of SF imaging, during which time all molecules in detectable states are pumped into dark vibrational levels $n_\text{loss}$ (which are high-lying and therefore neglected for the remainder of the simulation) at a rate corresponding to a branching ratio of $8.5\times10^{-5}$ and a scattering rate of $45 \times 10^3$ s$^{-1}$. Additionally, molecules in repumped vibrational levels are pumped back into $\widetilde{X}^2\Sigma^+(000)(N=1)$ with a 1 ms timescale during the imaging, and molecules in $\widetilde{X}^2\Sigma^+(000)(N=1, J=1/2)$ are pumped into $\widetilde{X}^2\Sigma^+(000)(N=1, J=3/2)$ as described above. All dynamics present in eqn. \ref{eqn:rateeqn} remain active throughout the imaging period.

After imaging, the rate equation propagation proceeds differently for the $\widetilde{X}(000)$, $\widetilde{X}(010)$, and $\widetilde{X}(100)$ states. For the $\widetilde{X}(000)$ data, the populations propagate according to eqn. \ref{eqn:rateeqn} for a variable time. For comparison to the experimental data, $t=0$ occurs immediately after the imaging light is turned off. The second SF image used to detect the molecules is not included in the model, since the number of imaged molecules is proportional to the number of detectable molecules at the start of the imaging, and the 50 ms imaging time is short compared to the lifetime of the $\widetilde{X}(000)$ state populated by the majority of imaged molecules. We have run the model including the second image in the rate equations and confirmed that this approximation negligibly changes the results.

\begin{figure}
	\centering
	\includegraphics[width= 0.6\textwidth]{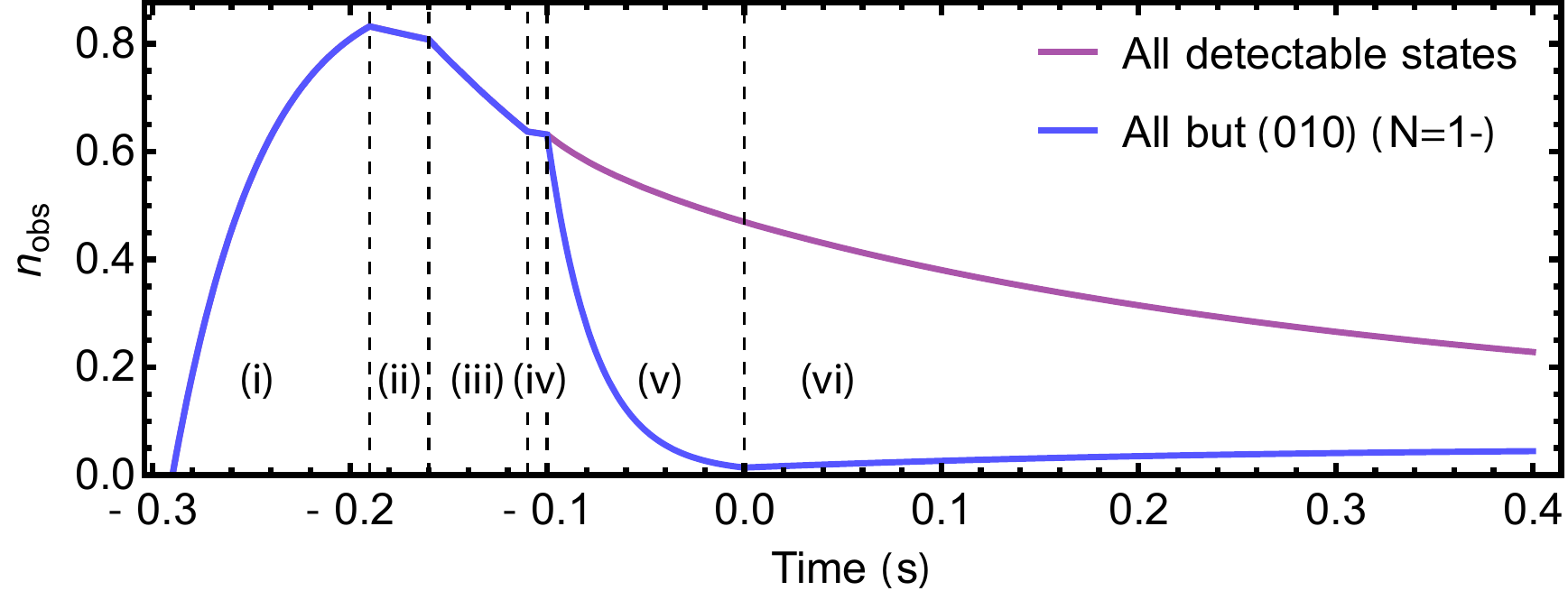}
	\caption{Population in detectable states (purple curve) and detectable states except $\widetilde{X}(010)(N=1^-)$ (blue curve) calculated from the best-fit rate equation for the $\widetilde{X}(010)$ lifetime sequence. The steps in the sequence are (i) ODT loading (ii) hold (iii) first image (iv) hold (v) optical pumping into $\widetilde{X}(010)$ and (vi) lifetime hold. $t=0$ here is the same as in Fig. 4b of the main text.}
	\label{fig:sequence}
\end{figure}

For the $\widetilde{X}(010)$ and $\widetilde{X}(100)$ data, the molecules are held in the ``dark'' (propagation according to eqn. \ref{eqn:rateeqn} only) for a short time after imaging (10 ms for $(010)$ and 90 ms for $(100)$). Molecules in detectable states are then pumped into the excited vibrational state at a rate $R_{\text{pump},i}=R_\text{scatt}v_iS_i$, where $R_\text{scatt} = 45 \times 10^3$ s$^{-1}$ is the SF scattering rate (see Sec. \ref{scattering_rate}), $v_{(010)} = 8.2 \times 10^{-4}$ is the vibrational branching ratio (VBR) to $\widetilde{X}(010)(N=1^-)$, and $v_{(100)} = 4.75 \times 10^{-2}$ is the VBR to $\widetilde{X}(100)(N=1)$. The rotational branching factors for $(010)$ are $S_{1/2}=0.73$ and $S_{3/2}=0.27$ to $J=1/2$ and $J=3/2$, respectively \cite{baum2021establishing}. For $(100)$ they are $S_{1/2}=2/3$ and $S_{3/2}=1/3$. There is also loss to dark states, $R_\text{dark} = R_\text{scatt}v_\text{dark}$, where $v_\text{dark} = 8.5 \times 10^{-5}$. The $\widetilde{X}(010)$ pumping occurs for 100 ms and the $\widetilde{X}(100)$ pumping occurs for 2 ms; in both cases the dynamics in eqn. \ref{eqn:rateeqn} remain active. After optical pumping, the populations propagate according to eqn. \ref{eqn:rateeqn} for a variable time. $t=0$ is defined as immediately after the optical pumping/state transfer light is turned off. The observable population, as well as the detectable population in all states besides $\widetilde{X}(010)(N=1^-)$, is plotted in Fig. \ref{fig:sequence} for the full $\widetilde{X}(010)$ lifetime sequence.

The model results are fit to the experimental data as follows. The rate equations for all three lifetime measurements are solved for a range of values of $|\text{d}\vec{\mu}/\text{d}Q_1|$, $|\text{d}\vec{\mu}/\text{d}Q_2|$, and $\tau_\text{vac}$, on a discrete grid surrounding the global minimum of the fit. The step size is 0.025 D for the dipole derivatives and 0.5 s for the vacuum lifetime. The observable population as a function of time, $n_\text{obs}(t) = \sum_{i\in\{\text{det}\}} n_i(t)$, is then calculated for comparison to the experimental data. Here $\{\text{det}\}$ is the subset of states which are detectable, i.e., the $(N=1, J=1/2-)$, $(N=1, J=3/2-)$, and $(N=2, J=3/2-)$ levels of repumped vibrational states. The results are then scaled by constant prefactors $a_{(000)}$, $a_{(010)}$, and $a_{(100)}$ for the three measurements, and a constant offset $a_{\text{off},i}$ (present due to imperfections in the imaging background subtraction, or imperfect state preparation) is added to each of the three traces. These scale factors are also scanned on a grid, with a step size of 0.05 for the amplitudes and 0.02 for the offset. The resulting fit functions depend on a total of 9 fit parameters:
\begin{align}
	n_{\text{fit},(000)}(t) &= a_{(000)}n_{\text{obs},(000)}(t,    \left|\frac{d\vec{\mu}}{dQ_1}\right|,
	\left|\frac{d\vec{\mu}}{dQ_2}\right|,
	\tau_\text{vac}) + a_{\text{off},(000)}, \\
	n_{\text{fit},(010)}(t) &= a_{(010)}n_{\text{obs},(010)}(t,    \left|\frac{d\vec{\mu}}{dQ_1}\right|,
	\left|\frac{d\vec{\mu}}{dQ_2}\right|,
	\tau_\text{vac}) + a_{\text{off},(010)}, \\
	n_{\text{fit},(100)}(t) &= a_{(100)}n_{\text{obs},(100)}(t,
	\left|\frac{d\vec{\mu}}{dQ_1}\right|,
	\left|\frac{d\vec{\mu}}{dQ_2}\right|,
	\tau_\text{vac}) + a_{\text{off},(100)}.
\end{align}
These are the functions plotted in Fig. 4 of the main text.

To perform the fit, the sum of squared errors,
\begin{equation}
	S = \sum_i \left[n_{i,(000)} - n_{\text{fit},(000)}(t_i)\right]^2 + \sum_i \left[n_{i,(010)} - n_{\text{fit},(010)}(t_i)\right]^2 + \sum_i \left[n_{i,(100)} - n_{\text{fit},(100)}(t_i)\right]^2,
\end{equation}
(where $t_i$ and $n_i$ are the time and measured survival from the experimental data) is calculated for each point on the grid described above. The optimal parameters are determined by fitting the 10,000 lowest $S$ values (approximately 1\% of the full grid) to a second-order polynomial. The fit is constrained to these points in order to minimize the effect of higher-order curvature of the error surface on the fit. The parameter errors determined from this fit are negligible compared to error sources described below.

\begin{figure}
	\centering
	\includegraphics[width=0.84\textwidth]{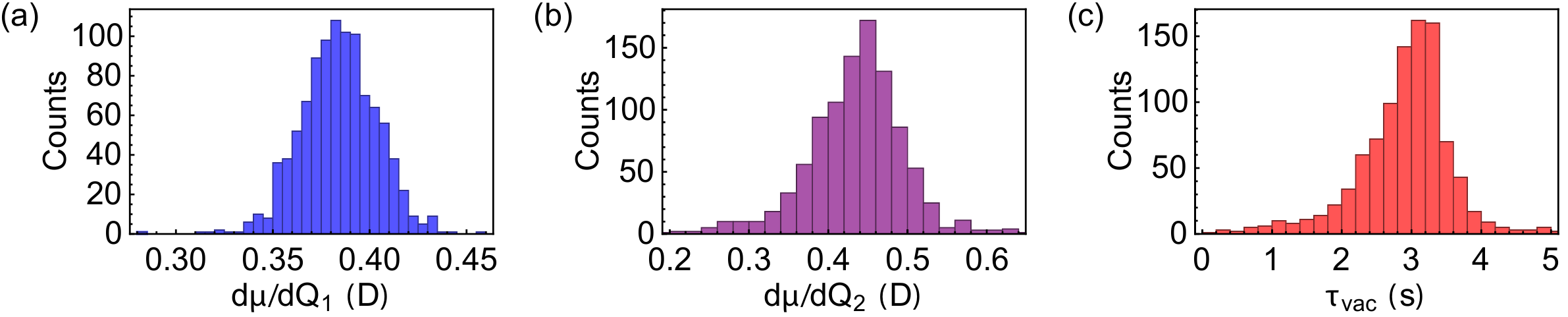}
	\caption{Histograms of the parameters (a) $|d\mu/dQ_1|$, (b) $|d\mu/dQ_2|$, and (c) $\tau_\text{vac}$ generated by fitting the rate equation model to $\sim$1000 synthetic data sets sampled according to the error of the measured data points.}
	\label{fig:histograms}
\end{figure}

The primary source of uncertainty in the fitted lifetimes is due to correlations between parameters, for example between the bending mode lifetime and fit amplitude, $|d\vec{\mu}/dQ_2|$ and $a_{(010)}$. These correlations have the potential to make the fit results sensitive to the precise values of the individual data points. Additionally, the $\widetilde{X}(010)$ radiative decay occurs on a similar timescale to blackbody excitation and vacuum loss, making it difficult to isolate. To account for these factors and adequately estimate the parameter errors, we repeat the fitting procedure $\sim$1000 times, each time fitting to a ``synthetic'' data set produced by sampling the value of each point from a normal distribution generated from the mean and standard deviation of the measured data point (i.e., the data plotted in Fig. 4 of the main text). This is meant to approximate the variation of the fit parameters expected if the experiment were to be repeated 1000 times. Histograms of the resulting fit parameters are shown in Fig. \ref{fig:histograms}. The parameters and error bars are determined by taking the median of the histogram as the parameter value and the middle 68\% of the distribution as the (typically asymmetric) $1\sigma$ confidence interval. Results are shown in Tab. \ref{tab:rateresults}.

\setlength{\tabcolsep}{0.5em} 
\begin{table}
	\centering
	\begin{tabular}{| c | c | c |}
		\hline
		Parameter & Median & 68\% Conf. Int.  \\
		\hline
		$|d\mu/dQ_1|$ & 0.384 D & (0.365, 0.404) D \\
		$|d\mu/dQ_2|$ & 0.440 D & (0.380, 0.486) D \\
		$\tau_\text{vac}$ & 2.99 s & (2.33, 3.43) s \\
		$a_{(000)}$ & 1.36 & (1.29, 1.43) \\
		$a_{(010)}$ & 1.31 & (1.18, 1.40) \\
		$a_{(100)}$ & 1.18 & (1.11, 1.24) \\
		$a_{\text{off},(000)}$ & -0.015 & (-0.044, 0.018) \\
		$a_{\text{off},(010)}$ & 0.045 & (0.032, 0.061) \\
		$a_{\text{off},(100)}$ & 0.063 & (0.049, 0.076) \\
		\hline
	\end{tabular}
	\caption{Fit parameters and 68\% confidence intervals for the rate equation fit.}
	\label{tab:rateresults}
\end{table}

The state lifetimes are calculated from the fitted parameters and eqns. \ref{eqn:spont}-\ref{eqn:BBR} by summing over all allowed transitions, i.e.,
\begin{align}
	\tau_{\text{spont},i} &= \left(\sum_{j<i} A_{ij}\right)^{-1}, \\
	\tau_{\text{bbr},i} &= \left(\sum_j R_{ij}\right)^{-1},  \\
	\tau_{\text{tot},i} &= \left(\frac{1}{\tau_{\text{spont},i}}+\frac{1}{\tau_{\text{bbr},i}}+\frac{1}{\tau_{\text{vac}}}\right)^{-1}.
\end{align}
Note that even though population will return to the state at later times as the vibrational distribution thermalizes, the coherence time is set by the lifetime given here since returning population is incoherent.

The amplitudes $a_{(000)}$, $a_{(010)}$, and $a_{(100)}$ can be understood as scale factors between the calculated populations, which are normalized to 1, and the measured populations, which are normalized to the number of molecules detected in the first image. From the fitted model, the detectable population at the start of the first image is $n_\text{obs}(t= t_\text{img}) = 0.81$, so the model results need to be scaled by $1/0.81 = 1.24$ to match the experimental data. The fitted amplitudes in Tab. \ref{tab:rateresults} are in good agreement with this expectation. 

\section{Calculations of lifetimes for the $X^2\Sigma^+(010)$ states of CaOH, SrOH, and YbOH} \label{ab_initio}

\subsection{Computational details}
The calculation of a spontaneous decay rate $\Gamma_\text{sp}$ (in a.u.)
\begin{equation}
	\Gamma_\text{sp} = \frac{4\omega^3  |\vec{\mu}|^2}{3c^3},
\end{equation}
involves the transition dipole moment $|\vec{\mu}|$ and the energy difference $\omega$ between two vibrational states.
Here $c$ is the speed of light.
The present discrete variable representation (DVR) \cite{Colbert1992, zhang2021accurate} calculations expand the vibrational wave functions in terms of real space basis functions on a grid with 21 evenly spaced points in the range $[-4.0Q,4.0Q]$ for bending modes and the metal-oxygen stretching mode as well as 28 points in $[-6.8Q,4.0Q]$ for the O-H stretching mode, in which $Q$ represents a dimensionless normal mode of the $X^2\Sigma^+$ state.
The transition dipole moment between $X^2\Sigma^+(010)$ and $X^2\Sigma^+(000)$ was calculated as 
\begin{equation}
	\label{expectation}
	\vec{\mu} = \langle \chi_{000}(\mathbf{R}) | \vec{\mu} (\mathbf{R}) | \chi_{010} (\mathbf{R})\rangle ,
\end{equation}
in which $\mathbf{R}$ represents the normal coordinates, $\chi_{000}(\mathbf{R})$ and $\chi_{010}(\mathbf{R})$ are the vibrational wave functions, and $\vec{\mu} (\mathbf{R})$ is the dipole moment function of the $X^2\Sigma^+$ state. 
The dipole moment functions were obtained by fitting calculated dipole moment values for structures near the equilibrium structure into a polynomial of normal coordinates. We fitted 625 dipole-moment values computed on a grid, which consists of 5 evenly spaced points in the range $[-0.2Q,0.2Q]$ for each normal mode, into a fourth-order polynomial function in terms of the normal coordinates.  
The dipole-moment calculations on this grid were performed using the same computational methods as those used for calculations of potential energy surfaces in Ref. \cite{zhang2021accurate}, i.e., the equation-of-motion electron-attachment coupled-cluster singles and doubles (EOMEA-CCSD) method \cite{EOMCC1Stanton1993,EOMEA_Nooijen95} and correlation-consistent quadruple-zeta (QZ) basis sets for CaOH and triple-zeta (TZ) basis sets for SrOH and YbOH. \cite{dunning1989gaussian,koput2002ab,de2001parallel,hill2017gaussian,lu2016correlation}
The detailed information about the basis sets, frozen orbitals, and the potential energy surfaces for DVR calculations have been documented in Ref. \cite{zhang2021accurate}.

The overall lifetime was obtained by further including contributions from the black-body radiation (BBR) induced transitions from the $X^2\Sigma^+(010)$ state to higher excited states at 300 K using the formulae developed in Ref. \cite{vanhaecke2007precision,beterov2009quasiclassical}. The BBR decay rate of the $i$ state can be evaluated by summing over other states $i^{\prime}$,
\begin{equation}
	\Gamma_\text{BBR} = \sum_{i^{\prime}}\Gamma_\text{sp}(i\to i^{\prime})\frac{1}{e^{\omega_{ii^{\prime}}/k_B T}-1},
\end{equation}
where $k_B$ is the Boltzmann constant and $\omega_{ii^{\prime}}$ is the energy difference between $i$ and $i^{\prime}$ states.
In the present calculations, $i$ corresponds to the $X^2\Sigma^+(010)$ state
and $\Gamma_\text{BBR}$ receives non-negligible contributions from transitions to the $X^2\Sigma^+(020)$ and $X^2\Sigma^+(110)$ states. 
The calculated $X^2\Sigma^+(000)-X^2\Sigma^+(010)$ transition dipole moments ($|\vec{\mu}|$), spontaneous lifetimes ($\tau_\text{sp}$), and overall lifetimes ($\tau$) at 300 K for the $X^2\Sigma^+(010)$ state of CaOH, SrOH, and YbOH are summarized in Tab. \ref{tab:casryb}.
For the CaOH molecule, we also calculated the spontaneous and overall lifetimes for the $X^2\Sigma^+(100)$ state and the overall lifetime for the ground $X^2\Sigma^+(000)$ state. The results are summarized in Tab. \ref{tab:caoh}.

\begin{table}
	\begin{center}
		\begin{tabular}{ccccccc}
			\hline \hline
			&~&  $|\vec{\mu}|$   &~& $\tau_\text{sp}$ &~& $\tau$ (300 K)\\
			\hline
			CaOH  &~& 0.284  &~& 876 &~& 409 \\
			SrOH  &~& 0.268  &~& 902 &~& 439 \\
			YbOH  &~& 0.305  &~& 1020 &~& 440 \\
			\hline \hline
		\end{tabular}
		\caption{The $X^2\Sigma^+(000)-X^2\Sigma^+(010)$ transition dipole moments (in debye), spontaneous lifetimes (in ms), and overall lifetimes at 300 K (in ms) for the $X^2\Sigma^+(010)$ states of CaOH, SrOH, and YbOH.}
		\label{tab:casryb}
	\end{center}
\end{table}

\begin{table}
	\begin{center}
		\begin{tabular}{ccccccc}
			\hline \hline
			&~&  $|\vec{\mu}|$   &~& $\tau_\text{sp}$ &~& $\tau$ (300 K)\\
			\hline
			$X^2\Sigma^+(100)$  &~& 0.295  &~& 161  &~& 141 \\
			$X^2\Sigma^+(000)$  &~& -      &~& -    &~& 1143 \\
			\hline \hline
		\end{tabular}
		\caption{The $X^2\Sigma^+(000)-X^2\Sigma^+(100)$ transition dipole moment (in debye), the spontaneous and overall lifetimes at 300 K (in ms) for the $X^2\Sigma^+(100)$ state of CaOH, and the overall lifetime for the $X^2\Sigma^+(000)$ states of CaOH.}
		\label{tab:caoh}
	\end{center}
\end{table}

We also calculated the transition dipole moments and spontaneous lifetimes for the $X^2\Sigma^+(010)$ and $X^2\Sigma^+(100)$ states of CaOH using the harmonic approximation.
The transition dipole moment between $v=1$ and $v=0$ states within the harmonic approximation, $\vec{\mu}_\text{H}$, can be evaluated as $\vec{\mu}_\text{H} = \frac{1}{\sqrt{2}} d\vec{\mu}_\text{ele}/dQ$. The dipole derivative $d\vec{\mu}_\text{ele}/dQ$ is obtained from the linear coefficients of the fitted dipole function.
A comparison between the DVR results (``Calc.''), the calculated results using harmonic approximation [``Calc. (H)''], and the experimental measurements (``Exp.'') is given in Tab. \ref{tab:harmonic_tab}. 
The anharmonic contributions reduce the magnitude of computed transition dipole moments and hence increase the
computed lifetimes.
The computed spontaneous lifetimes compare favorably with the measured ones and are within the uncertainty of the measured values.

\begin{table}
	\begin{center}
		\begin{tabular}{ccccccccccccc}
			\hline \hline
			Vibrational state & \multicolumn{2}{c}{$|\vec{\mu}|$} &~& \multicolumn{2}{c}{$|d\vec{\mu}_\text{ele}/dQ|$}  &~& \multicolumn{3}{c}{$\tau_\text{sp}$} &~& \multicolumn{2}{c}{$\tau$ (300 K)}\\
			& Calc. (H) & Calc. &~& Calc. & Exp. &~& Calc. (H) & Calc. & Exp. &~& Calc. & Exp.\\
			\hline
			CaOH $X^2(010)$ & 0.301 & 0.284 &~& 0.426 & 0.479 &~& 0.78 & 0.88 & 0.72 &~& 0.41  & 0.41 \\
			CaOH $X^2(100)$ & 0.262 & 0.295 &~& 0.370 & 0.394    &~& 0.20 & 0.16 & 0.19 &~& 0.14  & 0.15 \\
			CaOH $X^2(000)$ & -     & -     &~& -    & -       &~& -   & -   & -   &~& 1.14 & 1.28 \\
			\hline \hline
		\end{tabular}
		\caption{The calculated transition dipole moments $\vec{\mu}$ (in debye), and calculated and experimentally determined derivatives of electronic dipole moments $|d\vec{\mu}_\text{ele}/dQ|$ (in debye), spontaneous lifetimes (in ms), and overall lifetimes at 300 K (in s). ``(H)'' represents the results obtained using harmonic approximation.  Experimental results at room temperature are the same as the ``overall'' lifetimes in the main text, but with the vacuum loss removed.}
		\label{tab:harmonic_tab}
	\end{center}
\end{table}

\subsection{Benchmark studies}

To investigate the accuracy of the dipole moment function, 
we calculated the dipole moment function using Hartree-Fock (HF), coupled-cluster \cite{Crawford00,Bartlett07} singles and doubles (CCSD), and CCSD with a non-iterative triple [CCSD(T)] \cite{Purvis82,Scuseria88a,Raghavachari89}, and EOMEA-CCSD methods with QZ and 5Z basis sets. 
The calculated vibrational transition dipole moments, electronic dipole derivatives, spontaneous lifetimes, and overall lifetimes for the $X^2\Sigma^+(010)$ state of CaOH using these dipole moment functions are summarized in Tab. \ref{tab:dipole1}.
The EOM-CCSD/QZ results agree very well with the CCSD/QZ and CCSD(T)/QZ values. For example, the EOM-CCSD/QZ value for the transition dipole moment amounts to 0.284 debye, which is in close agreement with the CCSD/QZ value of 0.290 debye and the CCSD(T)/QZ value of 0.286 debye. The remaining electron-correlation contributions are expected to be small. 
The EOM-CCSD/QZ and EOM-CCSD/5Z results also agree with each other closely, indicating that
the remaining basis-set effects are small. 

\begin{table}
	\begin{center}
		\begin{tabular}{ccccccccc}
			\hline \hline
			Computational method    &~&  $|\vec{\mu}|$ &~&  $|d\vec{\mu}/dQ|$  &~& $\tau_\text{sp}$ &~& $\tau$ (300 K)\\
			\hline
			EOM-CCSD/QZ &~& 0.284 &~& 0.426 &~& 876 &~& 409 \\
			EOM-CCSD/5Z &~& 0.285 &~& 0.426 &~& 867 &~& 402 \\
			HF/QZ       &~& 0.294 &~& 0.437 &~& 813 &~& 372 \\
			CCSD/QZ     &~& 0.290 &~& 0.432 &~& 836 &~& 391 \\
			CCSD(T)/QZ  &~& 0.286 &~& 0.426 &~& 863 &~& 405 \\
			\hline \hline
		\end{tabular}
		\caption{Transition dipoles (in debye), dipole derivatives (in debye), spontaneous lifetime (in ms), and overall lifetime (in ms) of $X^2(010)$ state of CaOH molecule obtained from dipole surfaces calculated using different methods.}
		\label{tab:dipole1}
	\end{center}
\end{table}

\begin{table}
	\begin{center}
		\begin{tabular}{ccccccc}
			\hline \hline
			Fitting range/Fitting order    &~&  $|\vec{\mu}|$   &~& $\tau_\text{sp}$ &~& $\tau$ (300 K)\\
			\hline
			$[-0.2Q,0.2Q]$/4th &~& 0.284  &~& 876 &~& 409 \\
			$[-2.0Q,2.0Q]$/4th &~& 0.287  &~& 854 &~& 399 \\
			$[-2.0Q,2.0Q]$/6th &~& 0.287  &~& 856 &~& 398 \\
			\hline \hline
		\end{tabular}
		\caption{Transition dipole moments $|\vec{\mu}|$ (in debye), spontaneous lifetimes $\tau_\text{sp}$ (in ms), and overall lifetimes $\tau$ at 300 K (in ms) for the $X^2\Sigma^+(010)$ state of CaOH obtained from dipole moment functions fitted using the original and the enlarged data sets as well as with increased order of polynomial in the fitting.}
		\label{fit}
	\end{center}
\end{table}

We have examined the sensitivity of the computed results with respect to the grid points used to fit the dipole moment function.
By looking into the contributions to the expectation value in eqn. (\ref{expectation}), we found that the contributions are mainly from the DVR basis functions within the range of $[-2.0Q,2.0Q]$ for each normal mode.
To ensure an accurate representation of this range, we performed dipole-moment calculations of on a grid of 625 points consisting of 5 evenly spaced points in $[-2.0Q, 2.0Q]$ for each normal mode. These computed dipole-moment values were added to the original data set in the fitting. The computed transition dipole moments, spontaneous lifetimes, and overall lifetimes using thus fitted dipole moment functions are summarized in Tab. \ref{fit}. There is a 3\% decrease of the computed lifetimes when using the dipole-moment function fitted using the enlarged data set. Increasing the order of polynomial to the 6th-order does not change the computed results significantly.

We have also used the EOM-CCSDT/TZ potential energy surface in the DVR calculation. 
This calculation gives a transition dipole moment of 0.286 debye, slightly larger than the value of 0.284 debye obtained from calculations using the EOM-CCSD/TZ surfaces. Therefore, the remaining correlation effects on the potential energy surfaces play a minor role. We note that the computed spontaneous lifetime is proportional to the cubic power of energy difference between the vibrational states and thus is very sensitive to this parameter. Since our calculated value of 356 cm$^{-1}$ is in close agreement with the measured value of 353 cm$^{-1}$, the corresponding error in the lifetime calculation is also expected to be small. 

Based the sources of errors discussed above, we give an error estimate of around 10\% for the computed transition dipole moments. This corresponds to an uncertainly of around 20\% for the computed lifetimes. 

\section{Calculations of polarizability for CaOH}

\begin{table}
	\begin{center}
		\begin{tabular}{ccccccccc}
			\hline \hline
			&~&  $\alpha_{xx/yy}$ &~&  $\alpha_{zz}$  \\
			\hline
			HF/aug-cc-pVTZ-unc            &~& 196.6 &~& 116.6  \\
			CCSD/aug-cc-pVTZ-unc          &~& 165.8 &~& 114.7  \\
			CCSD/aug-cc-pVQZ-unc          &~& 166.3 &~& 115.6  \\    
			CCSD(T)/aug-cc-pVTZ-unc       &~& 164.5 &~& 116.6  \\
			\hline \hline
		\end{tabular}
		\caption{Computed static polarizability tensor (a.u.) of CaOH.}
		\label{tab:dipole2}
	\end{center}
\end{table}

The static polarizability tensor for CaOH has been calculated using analytic second derivative techniques for the CCSD and CCSD(T) methods \cite{Gauss97, Christiansen98} and uncontracted augmented correlation-consistent polarized valence triple-zeta and quadruple-zeta (aug-cc-pVTZ-unc and aug-cc-pVQZ-unc) basis sets. As shown in Tab. \ref{tab:dipole2}, the triples correction, i.e., the difference between CCSD(T)/aug-cc-pVTZ-unc and CCSD/aug-cc-pVTZ-unc results, is less than 2\% of the total value. The CCSD/aug-cc-pVQZ-unc and CCSD/aug-cc-pVTZ-unc results differ by less than 1\%.  
We thus estimate the uncertainty of the CCSD/aug-cc-pVTZ-unc results to be less than 10\%.

The dynamic polarizability tensor at the trapping frequency of 1064 nm has been calculated at the CCSD/aug-cc-pVTZ-unc level of theory. This gives a value of 234.6 a.u. for $\alpha_{xx}$ and $\alpha_{yy}$, and a value of 142.6 a.u. for $\alpha_{zz}$. Based on the calculations of static polarizability tensor in the previous paragraph, we also estimate the uncertainty of the computed dynamic polarizability tensor to be less than 10\%. With an ODT power of $13.3~\text{W}$ and trap waist of $\sim$$25~\mu\text{m}$, the calculated polarizabilities result in a trap depth of $\sim$$600~\mu\text{K}$.

\clearfmfn


\begin{thebibliography}{75}
\bibitem [{Sta()}]{StateNotation}%
	\BibitemOpen
	\href@noop {} {}\bibinfo {note} {{The notation $(v_1 v_2 v_3)$ refers to the
			vibrational quanta in the three vibrational normal modes of CaOH, namely the
			Ca-O stretch, Ca-O-H bend, and O-H stretch modes, respectively. $N$ is the
			rotational quantum number, $J$ is the total angular momentum excluding
			nuclear spin, and $F$ is the hyperfine quantum number. For electronic
			transitions, a prime ($'$) refers to the excited state and a double prime
			($''$) refers to the ground state. A sign following the label (e.g.,
			$N''=1^-$) refers to the parity of the state, and is specified explicitly in
			cases where parity may not otherwise be clear.}}\BibitemShut {Stop}%
\bibitem [{Qua()}]{QuantumNumberEll}%
	\BibitemOpen
	\href@noop {} {}\bibinfo {note} {{$\ell$ is the projection of the vibrational
			orbital angular momentum in the bending mode onto the internuclear axis of
			the molecule. In $\widetilde{X}^2\Sigma^+(010)$, $\ell$ can be $-1$ or
			$+1$.}}\BibitemShut {Stop}%
\end{thebibliography}
\end{document}